\documentclass[prb,twocolumn]{revtex4}
\usepackage{graphicx}
\usepackage{color}
\usepackage{amsmath}

\begin{document}

\def\K{{\bf{K}}}
\def\Q{{\bf{Q}}}
\def\X{{\bf{X}}}
\def\Gbar{\bar{G}}
\def\tk{\tilde{\bf{k}}}
\def\k{{\bf{k}}}
\def\q{{\bf{q}}}
\def\x{{\bf{x}}}
\def\y{{\bf{y}}}

\title{Thermodynamics of the Quantum Critical Point at Finite Doping in 
the 2D Hubbard Model: A Dynamical Cluster Approximation Study}
\author{K. Mikelsons$^{1,2}$, E. Khatami$^{1,2}$, D. Galanakis$^{1}$, 
A. Macridin$^{3}$, J. Moreno$^{1}$, and M. Jarrell$^{1}$}
\address{$^{1}$Department of Physics and Astronomy, Louisiana State University, Baton Rouge, Louisiana, 70803, USA\\
$^{2}$Department of Physics, University of Cincinnati, Cincinnati, OH, 45221, USA\\
$^{3}$Fermilab, P. O. Box 500, Batavia, Illinois, 60510, USA}


\begin{abstract}
We study the thermodynamics of the two-dimensional Hubbard model within 
the dynamical cluster approximation. We use continuous 
time quantum Monte Carlo as a cluster solver to avoid the systematic error 
which complicates the calculation of the entropy and potential energy 
(double occupancy).  We find that at a critical filling, there is a 
pronounced peak in the entropy divided by temperature, $S/T$, and in the 
normalized double occupancy as a function of doping.  At this filling, 
we find that specific heat divided by temperature, $C/T$, 
increases strongly with decreasing temperature 
and kinetic and potential energies vary like $T^2\ln T$. 
These are all characteristics of quantum critical behavior. 
\end{abstract}

\maketitle

\paragraph*{Introduction-} The properties of the hole-doped 
cuprate phase diagram, including a pseudogap (PG) at low doping and 
unusual metallic behavior at higher doping, have led many investigators 
to propose that there is a quantum critical point (QCP) in the cuprate 
phase diagram at optimal doping~\cite{sachdev_92}. Some of the most 
compelling evidence for the QCP is from various thermodynamic 
experiments~\cite{tallon}.

In the previous work~\cite{raja} employing the dynamical cluster quantum 
Monte Carlo (QMC) method~\cite{hettler:dca, jarrell:dca} to calculate the
quasiparticle fraction for the two-dimensional (2D) Hubbard model, we 
found evidence for a QCP separating a region of Fermi liquid (FL) 
character at high doping from a region of non-Fermi liquid PG character 
at low doping.  At the critical doping, we found marginal Fermi liquid 
(MFL) character~\cite{Varma} which is also seen above the FL and PG 
temperatures.

In this paper, we provide further evidence for a QCP in the thermodynamic 
properties, including the energies, the specific heat and entropy.  At the 
critical filling, both the kinetic and potential energy data show a $T^2\ln T$ 
low-temperature dependence, leading to a $T\ln T$ specific heat.  The entropy 
divided by temperature, $S/T$, is strongly peaked at the critical point.  In 
contrast to the results of the previous study~\cite{raja}, these quantities 
are independent of the location and character of the Fermi surface and thus 
avoid ambiguity in their interpretation.

\paragraph*{Formalism-}
\label{sec:formalism}
We start with the 2D Hubbard Hamiltonian
\begin{equation}
H=H_k+H_p=\sum_{\k\sigma}\epsilon_{\k}^{0}c_{{\k}\sigma}^{\dagger}c_{{\k}\sigma}^{\phantom{\dagger}}+U\sum_{i}n_{{i}\uparrow}n_{{i}\downarrow} \,,
\label{eq:hubbard}\end{equation}
 where $c_{{\k}\sigma}^{\dagger}(c_{{\k}\sigma})$ is the creation (annihilation)
operator for electrons of wavevector ${\k}$ and spin $\sigma$,  
$n_{i\sigma} =c_{i\sigma}^{\dagger}c_{i\sigma}$ is the number operator,
$\epsilon_{\k}^{0}=-2t\left(\cos(k_{x})+\cos(k_{y})-2\right)$
with $t$ being the hopping amplitude between nearest neighbor sites, 
and $U$ is the on-site Coulomb repulsion. The primary focus of this Rapid Communication
is the behavior of the energies.  For lack of a better term, we will 
refer to the quadratic part of $H_k$ as the kinetic energy and the interaction part
as the potential energy. The corresponding energies per site are~\cite{FetterandWalecka}
\begin{eqnarray}
E_{k}&=&\frac{\left\langle H_k\right\rangle}{N} =\frac{T}{N}\sum_{\omega_{n},\k,\sigma}\epsilon_{\k}^{0}G_{\sigma}(\k,i\omega_{n})\label{eq:KE}\\
E_{p}&=&\frac{\left\langle H_p\right\rangle}{N} =\frac{T}{2N}\sum_{\omega_{n},\k,\sigma}\Sigma_{\sigma}(\k,i\omega_{n})G_{\sigma}(\k,i\omega_{n})\,, \label{eq:PE}
\end{eqnarray}
where $G_{\sigma}(\k,i\omega_{n})$ and $\Sigma_{\sigma}(\k,i\omega_{n})$ are the single-particle
Green function and self-energy respectively, with $\omega_{n}=(2n+1)\pi T$
and $N$ represents the number of sites.

Aside from the numerical calculations, we obtained analytically the leading 
low-$T$ behavior for $E_k$ and $E_p$ for each of the regions around the QCP
starting from the corresponding forms for the self-energy.
In the FL region, the
imaginary part of the self-energy has the form 
\begin{equation}
\Sigma_{FL}^{\prime\prime}\left(\omega\right)=-\alpha_{FL}\max\left(\omega^{2},T^{2}\right),\left|\omega\right|<\omega_{X},\label{eq:SigmaFL}
\end{equation}
 where $\omega_{X}$ is a cutoff frequency and $\alpha_{FL}>0$. In the
MFL region, this quantity becomes~\cite{Varma} 
\begin{equation}
\Sigma_{MFL}^{\prime\prime}\left(\omega\right)=-\alpha_{MFL}\max\left(\left|\omega\right|,T\right),\left|\omega\right|<\omega_{c}\,\label{eq:SigmaMFL}
\end{equation}
 where $\alpha_{MFL}>0$, and $\omega_{c}$ is a frequency cutoff. Finally
in the PG region, we consider the following Ansatz~\cite{k_yang_06}
for the imaginary part of the self-energy: 
\begin{equation}
\Sigma_{PG}^{\prime\prime}\left(\omega\right)=-\pi\Delta^{2}({\bf k})\delta\left(\omega-\epsilon_{{\bf k}}^{0}\right),\left|\omega\right|<\omega_{c}^{*},\label{eq:SigmaPG}
\end{equation}
where $\omega_{c}^{*}$ is a cutoff frequency and
$\Delta\left({\bf k}\right)=\Delta_{0}\left(\cos k_{x}-\cos k_{y}\right)$
with the constant $\Delta_{0}$ being the PG magnitude.

Inserting the forms~(\ref{eq:SigmaFL}), (\ref{eq:SigmaMFL}) and (\ref{eq:SigmaPG}) 
in (\ref{eq:KE}) and (\ref{eq:PE}) and performing a low-$T$ expansion in the resulting 
integrals, we obtained that both $E_p$ and $E_k$ exhibit a leading $T^{2}$ low 
temperature behavior in the PG and FL regions. Both energies display a $T^{2}\ln T$ 
behavior consistent with a $T^{2}\ln T$ total energy \cite{crisan} in the 
MFL region.

\paragraph*{Methodology-}
We solve the Hubbard model within the dynamical cluster approximation 
(DCA)~\cite{hettler:dca}.
The DCA is 
a cluster mean-field theory which maps the original lattice model 
onto a periodic cluster of size $N_c=L_c^2$ embedded in a self-consistent 
host. Spatial correlations up to a range $L_c$ are treated explicitly, 
while those at longer length scales are described at the mean-field level.   
However the correlations in time, essential for quantum criticality,  are 
treated explicitly for all cluster sizes.  To solve the cluster problem 
we use weak coupling expansion continuous time QMC (CTQMC)
method~\cite{rombouts99,rubtsov05} with highly optimized blocked and 
delayed updates~\cite{karlis}.
In the previous work~\cite{raja}, a Hirsch-Fye QMC (HFQMC) algorithm was 
used.  However, the Trotter error intrinsic to HFQMC, which is particularly 
large for the potential energy (double occupancy), prevented us from
calculating $S$ since systematic errors from multiple temperatures tend to 
accumulate~\cite{a_dare_07}. The CTQMC algorithm eliminates the systematic 
error, and generally has a lower overall computational cost than HFQMC, 
making it a better choice for these calculations. 
Unless otherwise
displayed, the statistical error bars in these calculations are smaller 
than the plotting symbols or line widths used.

Here, we study the normal state on a $N_c = 4\times4$ site cluster for $U=6t$.
We confirmed that increasing the cluster 
size to $N_c=24$ sites does not significantly alter our results.
We obtain the energies from Eqs.~(\ref{eq:KE}) 
and (\ref{eq:PE}), and, following Ref.~\onlinecite{f_werner_05}, the entropy from
\begin{equation}
S(\beta,n) = S(0,n) + \beta E(\beta,n) - \int_0^{\beta} E(\beta', n) d \beta' \,, 
\label{eq:S}
\end{equation}
where $S(0,n)=- n \ln\frac{n}{2} - (2-n)\ln\left(1-\frac{n}{2}\right)$, 
$n$ is the filling and $\beta=1/T$. Finally, since the 
DCA preserves thermodynamic consistency~\cite{maier:rev}, 
our results obey the Maxwell relation
\begin{eqnarray}
\left( \frac{ \partial S}{\partial n} \right)_{T,U} &=&  
-\left( \frac{ \partial \mu }{\partial T} \right)_{U,n} \,,
\label{eq:smu}
\end{eqnarray}
where $\mu$ is the chemical potential.

\paragraph*{Results-} 

\begin{figure}[t]
\begin{center}
\includegraphics*[width=3.3in]{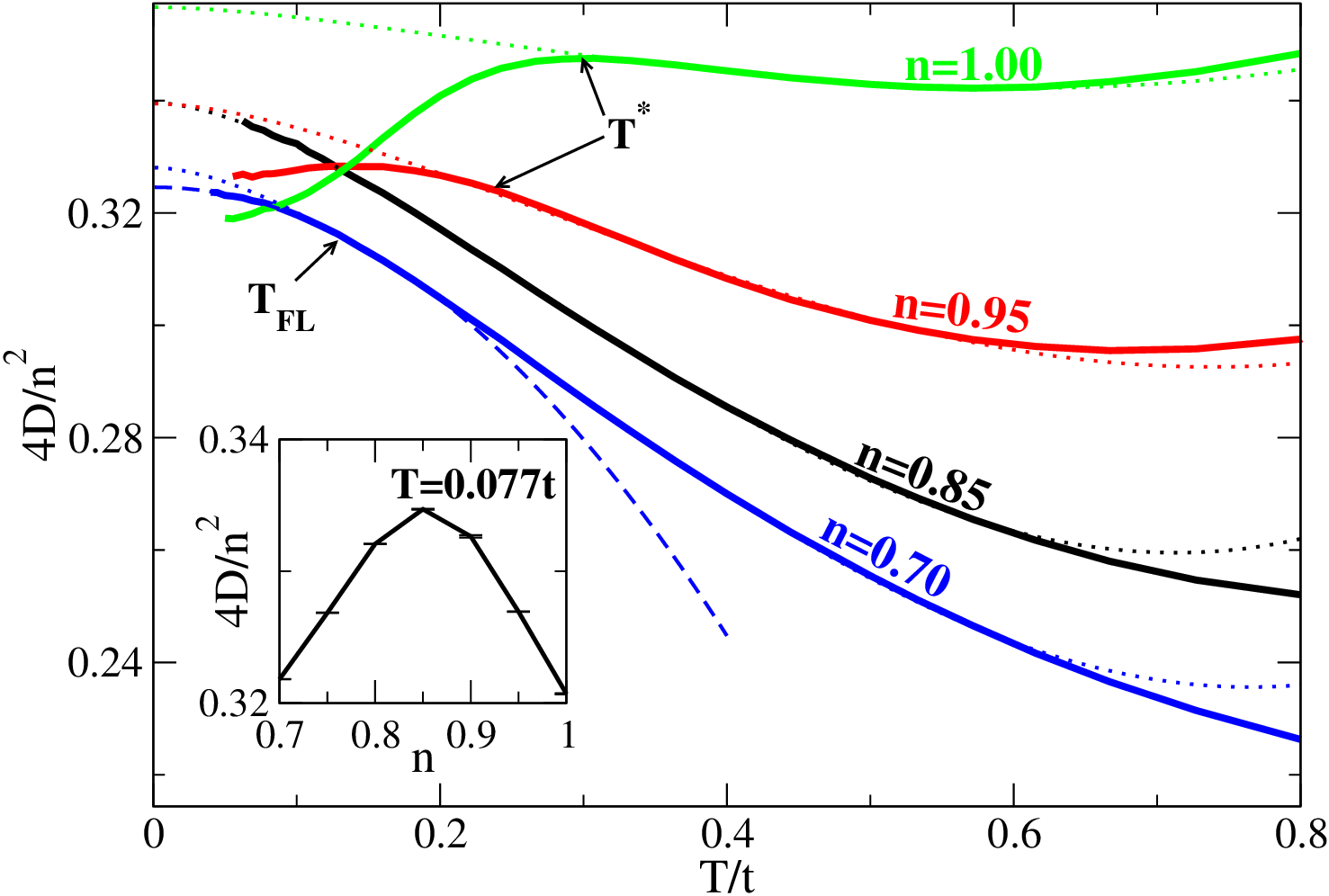}
\caption{(color online) Temperature dependence of normalized double 
occupancy $4D/n^2$. For a range of temperatures and values of filling, 
$D(T)$ fits well the MFL form (dotted lines): 
$D(T) = D_0 + \alpha T^2 \ln(T/\omega_c)$
 ($D_0$, $\alpha$ and $\omega_c$ are fitting parameters). 
At low temperatures, $D(T)$ shows FL behavior (dashed line) for $n=0.70$ 
($D(T) = D_0 + \alpha T^2$, $\alpha < 0$) and PG behavior for 
$n>0.85$. Indicated by arrows is the crossover temperature $T^*=0.13t$ for $n=0.70$ along
with the crossover temperatures $T_{FL}=0.24t$ and $T_{FL}=0.3t$ for  $n=0.95$ and $n=1.00$ respectively. 
For $n=0.85$, marginal 
Fermi liquid behavior persists down to lowest temperatures accessible 
with QMC. Inset: Normalized double occupancy versus filling for $T=0.077t$, 
showing a peak at the critical filling $n=0.85$. }
\label{fig:potential_energy}
\end{center}
\end{figure}

Fig.~\ref{fig:potential_energy} shows the normalized double occupancy, 
$4D/n^2$, as a function of temperature at fillings $n=1.00,\ 0.95,\ 0.85$ 
and $0.70$ together with fits to the MFL form for $n=0.95,\ 0.85$ and 
$0.70$, and the FL form for $n=0.70$. 
For $n=0.95$ and in a low temperature 
window, the $E_{p}$ first increases with decreasing temperature and then 
reaches a maximum at a temperature which coincides with the PG temperature 
($T^*$) found in a previous study~\cite{raja}. Further decreasing the 
temperature into the PG phase, $E_{p}$ deviates from the MFL character and 
begins to decrease. This indicates that close to half-filling, where the 
interactions are more relevant, the system tries to gain energy by lowering 
the $E_{p}$. A more pronounced decrease in the $E_{p}$ can be seen below the 
PG temperature in the half-filled case. At the critical filling, $n=0.85$, 
$E_{p}$ fits the MFL analytic form very well down to the lowest 
temperature reached. At this temperature, we find that the normalized double 
occupancy is maximal at the critical filling (see the inset of 
Fig.~\ref{fig:potential_energy}).  For $n=0.70$, the $E_{p}$ increases with decreasing temperature as in the MFL 
region. However, we find that below the FL characteristic temperature 
($T_{FL}$) (see Ref.~\onlinecite{raja}), it deviates from the MFL form and fits better to a 
quadratic function in $T$. Note that at intermediate temperatures 
($T^*$ or $T_{FL}<T<0.6t$), the MFL form fits the $E_{p}$ very well 
in all three doping regions.

\begin{figure}[t]
\begin{center}
\includegraphics*[width=3.3in]{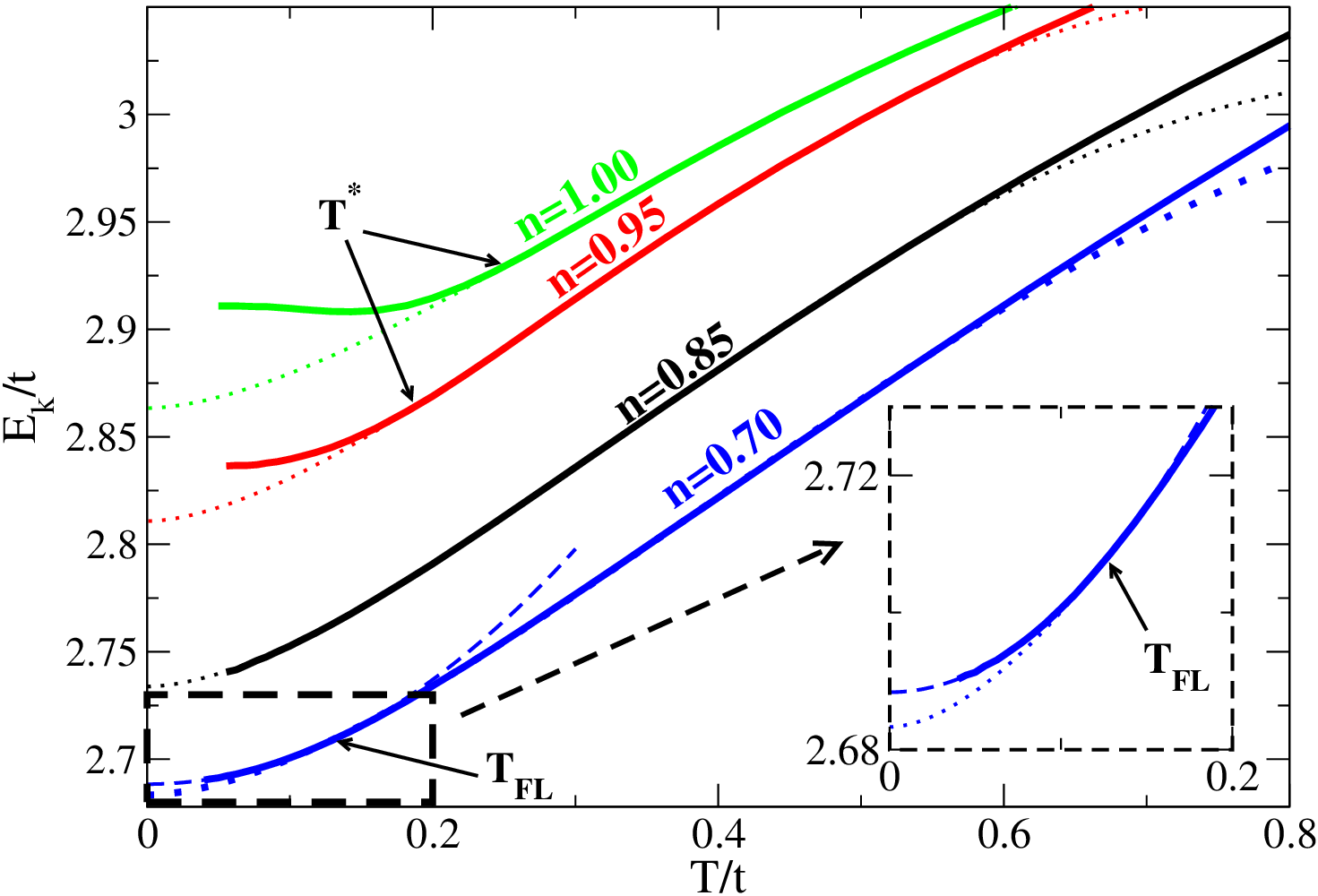}
\caption{(color online) Temperature dependence of kinetic energy. MFL form 
fits data for a range of intermediate temperatures (dashed lines). For 
$n=0.70$, crossover to FL behavior is indicated as $T_{FL}$, while for 
$n=0.95$ and $n=1.00$, crossover to PG region at $T^*$ is evident 
by strong departure of kinetic energy from MFL fit, which for $n=1.00$ even 
results in noticeable increase of kinetic energy with lowering temperature.}
\label{fig:kinetic_energy}
\end{center}
\end{figure}

In Fig.~\ref{fig:kinetic_energy}, we show $E_{k}$ versus temperature at the same 
 fillings. $T_{FL}$ and $T^*$ in Fig.~\ref{fig:kinetic_energy} represent 
the onset of FL and PG regions obtained from $E_{k}$ fits, and they have  the 
same value as those shown in Fig.~\ref{fig:potential_energy} for the $E_{p}$.
At low temperatures, while the $E_{k}$ deviates from its MFL fit for $n=1.00,\ 0.95$ 
and $0.70$, one sees no sign of a characteristic temperature scale at the 
critical filling.

We note that the low-$T$ behavior of both $E_{p}$ and $E_{k}$ follows
the analytical forms in the FL and MFL regions but slightly deviates from the 
predicted $T^2$ form in the PG region at low $T$. Also the 
energy scales below which the data deviate from the MFL $T^2\log T$ behavior
vanishes at a critical filling, $n_c=0.85$, where it persists down to the lowest 
temperature accessible, consistent with the existence of a QCP between 
FL and PG regions.

\begin{figure}[t]
\begin{center}
\includegraphics*[width=3.3in]{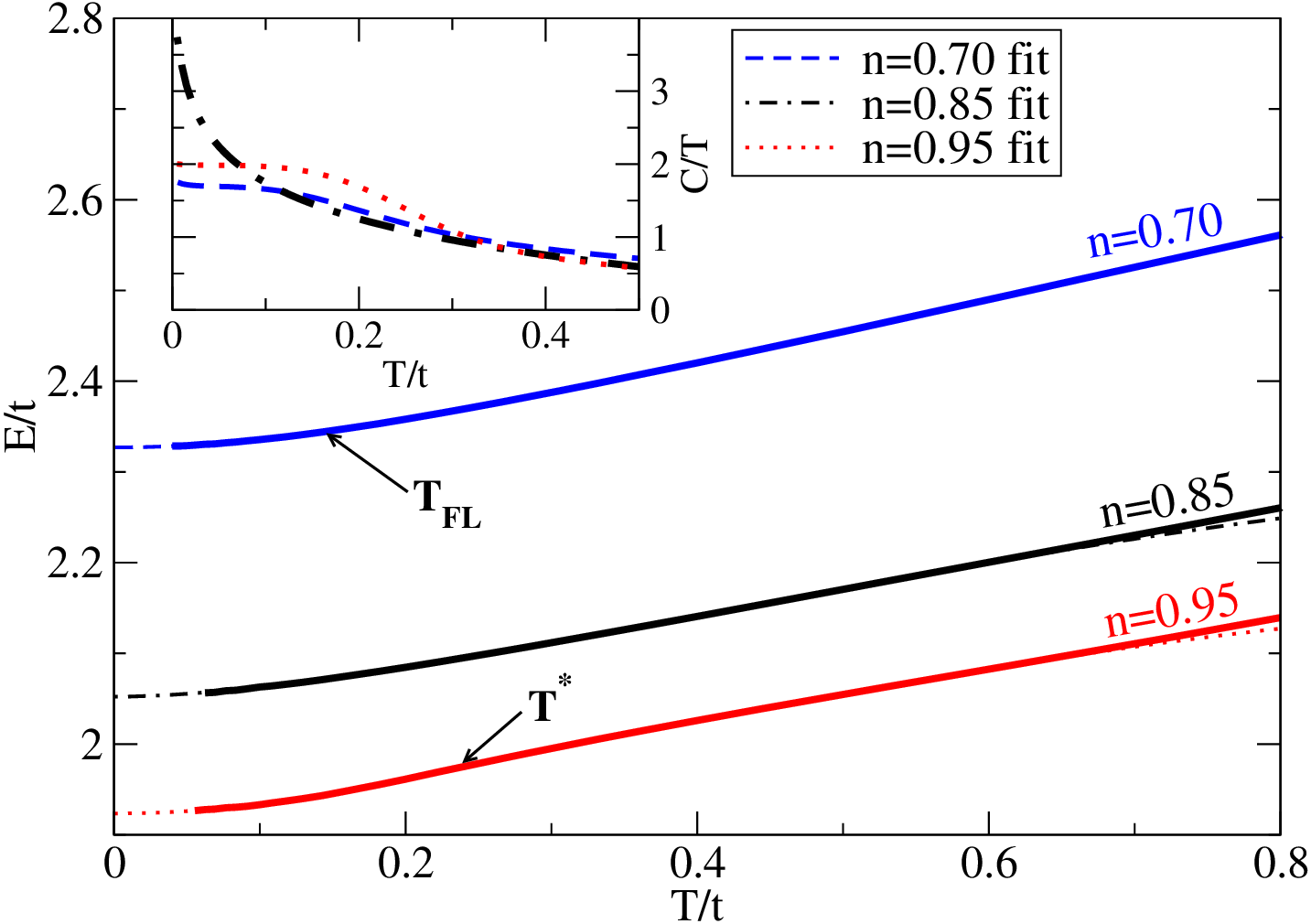}
\caption{(color online) Total energy per site $E$ versus temperature for different
fillings.  The data are fit to a crossover form of the energy, Eq.~(\ref{eq:Etot}) 
(dashed lines).  The values of $T_X$ determined from
the fit are indicated as $T_{FL}$ for $n=0.70$ and $T^*$ for $n=0.95$.
In the inset, the specific heat calculated from the fit is plotted 
versus temperature. }
\label{fig:Etot}
\end{center}
\end{figure}

The kinetic and potential energy data, together with the analytic forms
of Eqs.~(\ref{eq:PE}) and (\ref{eq:KE}) suggest that the total energy
 away from the half filling may be fit to the form
\begin{equation}
E(T) = E(0) + A f(T) T^2 + B\left(1-f(T) \right) T^2\ln\frac{T}{\Omega}\,,
\label{eq:Etot}
\end{equation}
where $f(T) = 1/\left( \exp\left( (T-T_X)/\theta\right) +1 \right)$ describes
the crossover from the MFL to the quadratic behavior, characteristic 
of a FL or presumably PG region.  $A$, $B$, $\theta$, $T_X$ and $\Omega$ are
the fitting parameters of the QMC energy data, as shown in Fig.~\ref{fig:Etot}.
At low $T$, the fit is indistinguishable from the data for all fillings.
The values of $T_X$ determined from the fit are indicated as $T_{FL}\approx 0.15t$ 
for $n=0.70$ and $T^*\approx0.24t$ for $n=0.95$ in agreement with the 
$E_{p}$ and $E_{k}$ fits. The calculation of the specific heat
is a notoriously difficult problem and usually involves a fit of $E(T)$ to
a regularized (smooth) functional form~\cite{carey_C,Andy_C}.  Here, we
already have an excellent fit, so we obtain $C/T$ from a derivative of 
the fit divided by temperature.  The result is shown in the inset.  
As expected for $n=0.70$, at low temperatures, $C/T$ is flat in $T$ 
consistent with FL formation.  This behavior is also seen for the data 
in the PG region, $n=0.95$, but the data at the critical filling
$n=0.85$, shows a weak divergence at low $T$ consistent with quantum
critical behavior.

\begin{figure}[t]
\begin{center}
\includegraphics*[width=3.3in]{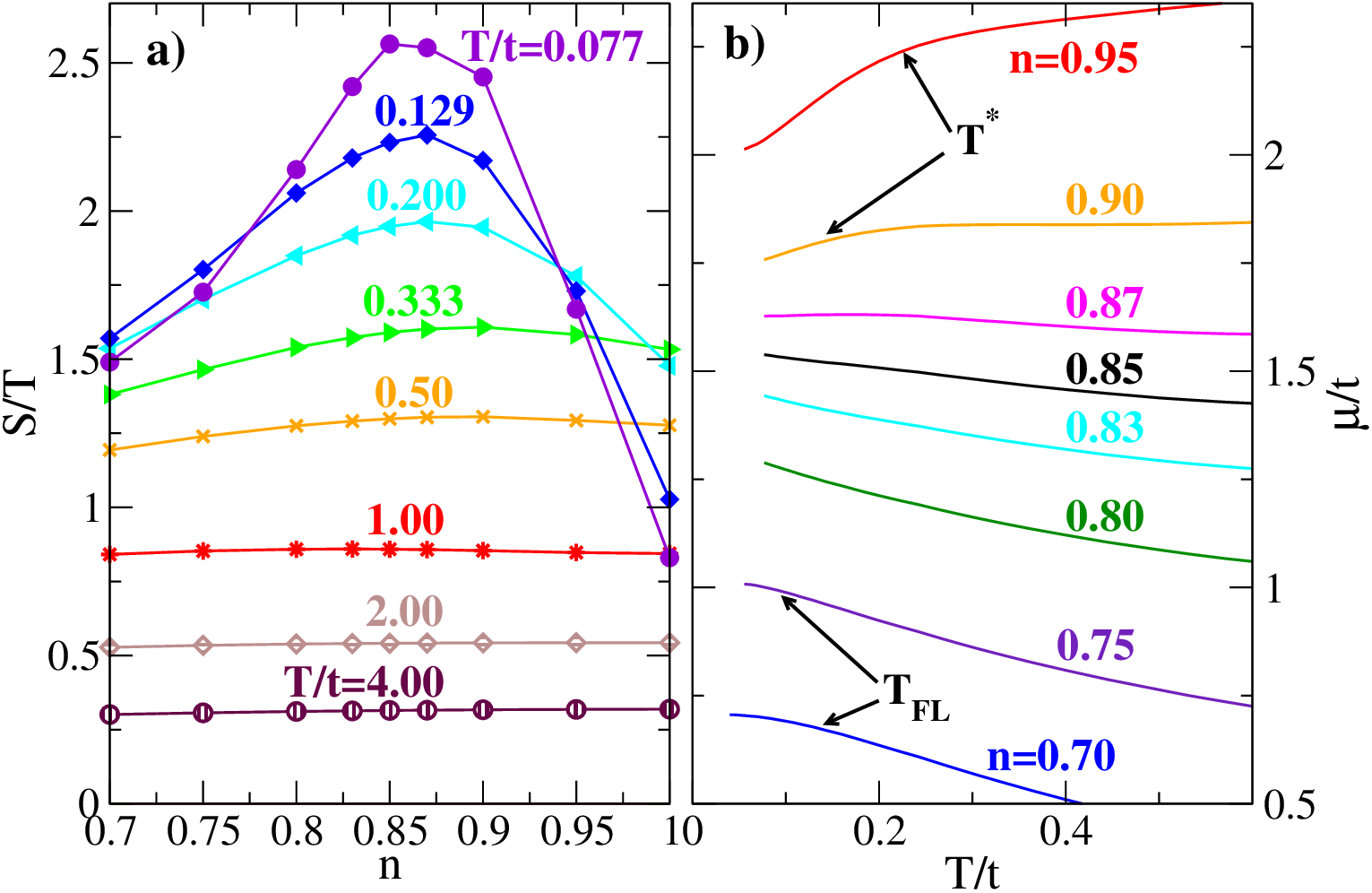}
\caption{(color online) Left panel: $S/T$ filling dependence showing 
emergence of a peak at $n=0.85$ at low temperatures. Right panel: 
Chemical potential temperature dependence for a range of fillings 
with PG and FL energy scales shown as $T^*$ and 
$T_{FL}$ for $n=0.95$ and $n=0.70$, respectively. Note that the 
position of the maximum of entropy in the left panel corresponds 
to $\partial \mu / \partial T = 0$ in the right panel. As temperature 
is lowered, the maximum of entropy shifts towards lower $n$, 
causing local particle-hole symmetry for $n=0.85$ at low $T$ (see text).}
\label{fig:S_vs_fill}
\end{center}
\end{figure}

The entropy per site exhibits intriguing behavior near the critical 
filling as the system is cooled, confirming much of the physics seen
in $C/T$ without the need for a fit or a numerical derivative.  It is 
more strongly quenched in the FL and PG regions than in the MFL region 
with decreasing $T$, resulting in a maximum in $S/T$ versus filling 
at $n=0.85$ at the lowest accessible temperature (see 
Fig.~\ref{fig:S_vs_fill} (a)).  The fact that $S/T$ continues to rise 
at the critical doping as $T\to 0$ is consistent with the increase in 
$C/T$. Similarly, the low temperature $S/T$ curves at different
temperatures nearly overlap for $n<0.85$ in agreement with a constant 
$C/T$ and indicative of a FL.  One can see the strong decrease in 
$S/T$ with decreasing temperature when the system crosses over to the 
PG phase around $T^*$ at large fillings (e.g. at half-filling).   In 
fact in the PG region, $n>0.85$, $S/T$ coincide for different 
temperatures only at one filling, roughly $n=0.95$. The flat low-T 
$C/T$ seen at this filling is accidental, and for higher filling one 
would expect $C/T$ to be strongly suppressed at low $T$
\footnote{This behavior in $S(T)$ at $n=1$ is likely an artifact of 
the DCA used here where the self energy is not interpolated.  An 
interpolated self energy, as used in e.g., Ref.~[\onlinecite{raja}], would preserve
the d-wave character of the PG and the $T^2$ character of $E(T)$.}.

According to Eq.~(\ref{eq:smu}), a local maximum in $S/T$ versus $n$ 
corresponds to a flat chemical potential as a function of temperature.
Therefore, at low $T$, the critical filling can be identified from the
temperature dependence of $\mu$ for different fillings.  This is shown 
in Fig.~\ref{fig:S_vs_fill} (b) 
where, one can see that the near temperature independence of $\mu$ 
at $n=0.90$ for $0.25t<T<0.50t$ evolves  
into  a broad maximum centered around $T=0.15t$ for $n=0.87$ which presumably 
moves to $n=0.85$ at low enough temperatures. These 
observations are consistent with the evolution of the maximum in $S/T$ 
versus $n$ as the temperature is lowered from $0.50t$ to $0.08t$ (see
Fig.~\ref{fig:S_vs_fill} (a)). In analogy to the half-filled case, a 
stationary chemical potential can be the signature of local particle-hole 
symmetry. This is consistent with the observation of near particle-hole  
symmetry in the cuprates in the proximity of optimal doping~\cite{s_chakraborty_08}.
It is also in agreement with previous 
results showing that at $n\approx 0.85$ and for the same model parameters, 
the low energy density of states displays particle-hole 
symmetry~\cite{raja}.

\paragraph*{Discussion-}  

The results presented here strongly favor an interpretation involving
a QCP as opposed to a simple crossover from a
FL to a non-Fermi liquid as the filling increases towards one.   The 
$T^2\ln T$ behavior of the kinetic and potential energies, the peak 
in $S/T$ which sharpens as $T$ falls, and the logarithmic behavior of 
the specific heat $C/T \sim \ln T$ at $n=0.85$ together with the wide 
range of temperatures and fillings influenced by the critical
point are clear signatures of quantum criticality. 

The peak in the low temperature normalized double occupancy observed at 
the critical filling is also interesting.  Near half filling $4D/n^2$ is 
suppressed by strong correlations. As the temperature is lowered 
$(T<T^*)$, it is suppressed further to allow the system to gain the most 
magnetic exchange energy.  As the filling decreases towards zero, the 
Fermi energy will become smaller than $U$, and again $4D/n^2$ is suppressed.  
So for any finite $U$ we expect a peak in $4D/n^2$ at some finite doping.  

At the critical filling, in addition to the enhanced $C/T$, we find a 
peak in the charge susceptibility (not shown).  These charge fluctuations 
may be responsible for the location of the peak $4D/n^2$ near the critical
doping. The peak in the charge susceptibility becomes sharper as the 
temperature is lowered, or when we include a next-near neighbor hopping 
$t'>0$.  This behavior suggests the possibility that the QCP may be 
associated with a charge instability seen previously~\cite{alex_PS}.   
This topic, as well as relation of the QPC and charge fluctuations to 
superconductivity, will be explored in an upcoming publication.

\paragraph*{Conclusion-}
We study the thermodynamics of the 2D Hubbard model using the dynamical
cluster approximation with CTQMC as a cluster solver.  The
latter eliminates the Trotter error which complicates the 
calculation of the entropy and potential energy.  At the critical doping 
we find that $C/T$, obtained from a fit of $E(T)$,  
increases strongly with decreasing temperature and the kinetic 
and potential energies are consistent with $T^2\ln T$.   
Near the critical filling, we find a pronounced peak in $S/T$ which grows
as the temperature falls, consistent with the growth of $C/T$.  We also
find a peak in the normalized double occupancy as a function of doping.   
These are all characteristics of quantum criticality. 

\paragraph*{Acknowledgments-}   We would like to thank 
P.~Phillips,
S.~Kivelson,
D.~J.~Scalapino,
A.~M.~Tremblay,
and
C.~Varma,
for useful conversations.  This research was supported by 
NSF DMR-0706379. JM and MJ are also supported by the NSF PIRE project OISE-0730290. 
This research used resources of the National Center for 
Computational Sciences at Oak Ridge National Laboratory, which is supported 
by the Office of Science of the U.S. Department of Energy under Contract 
No. DE-AC05-00OR22725.

\end{document}